\documentstyle[ltwol]{article}

\def\be{\begin{equation}}
\def\ee{\end{equation}}
\def\bea{\begin{eqnarray}}
\def\eea{\end{eqnarray}}

\input psfig.sty

\begin{document}

\title{
  Width difference of the $B_s$ mesons from lattice QCD
  }

\author{
  Shoji Hashimoto and Norikazu Yamada\\
  for the JLQCD collaboration
  }

\address{
  High Energy Accelerator Research Organization (KEK),
  Tsukuba 305-0801, Japan\\
  }


\twocolumn[\maketitle\abstracts{
  We extend our previous studies to calculate the $B$ meson $B$
  parameters $B_B$ and $B_S$ on the lattice, and present an update of
  the results for the $B_s$ meson width difference.
  We perform an extensive study of systematic errors in the quenched
  calculation of the $B$-parameters, and find that the systematic
  errors are in good control using the NRQCD action for heavy quark.
  We also report our preliminary results from unquenched simulations.
}]

\section{Width difference $\Delta\Gamma_{B_s}$}
The width difference $\Delta\Gamma_{B_s}$ in the $B_s-\bar{B}_s$
system opens a new possibility to measure the CKM angles and to search
for new physics, once it is measured and found to be sizable
\cite{Dunietz:2000cr}. 
Theoretical prediction of $\Delta\Gamma_s$ is, therefore, desirable,
and the most reliable calculation has been obtained using the Heavy
Quark Expansion under an assumption of the quark-hadron duality
\cite{Beneke:1996gn,Beneke:1999sy}, and a recent summary is given in
\cite{Beneke:2000cu}.
In this calculation, nonperturbative inputs are necessary for the $B$
meson $B$ parameters $B_B$ and $B_S$, for which the lattice QCD may
provide first-principles calculation starting from the QCD
lagrangian. 
In this report, we present an update of the lattice calculation of
$B_B$ and $B_S$ by the JLQCD collaboration.
Our previous works are published in 
\cite{Hashimoto:2000yh,Hashimoto:2000eh,Yamada:2001ym}.

The main contribution to $\Delta\Gamma_{B_s}$ comes from the
$c\bar{c}$ final state, to which both $B_s$ and $\bar{B}_s$ can decay,
and the theoretical expression obtained using the Heavy Quark
Expansion  
\footnote{
  As pointed out by Falk at the conference, the operator product
  expansion could be dangerous for this quantity, because the energy
  release is not large enough for the $c\bar{c}$ final states. 
  The quark-hadron duality could also be questionable, as the final
  states are saturated by a limited number of exclusive modes.
  Unfortunately, no rigorous method is known to estimate the violation
  of these assumptions.
}
was given by Beneke \textit{et al.} as 
\cite{Beneke:1996gn,Beneke:1999sy}
\begin{eqnarray}
  \lefteqn{
    \Delta\Gamma_{B_s} =
    \frac{G_F^2 m_b^2}{12\pi M_{B_s}}
    |V_{cb}^* V_{cs}|^2
    }
  \nonumber\\
  & &
  \times
  \left[
    c_L(z) \langle{\cal O}_L\rangle
    +
    c_S(z) \langle{\cal O}_S\rangle
    + c_{1/m}(z) \delta_{1/m}
  \right],
\end{eqnarray}
where $c_L(z)$, $c_S(z)$ and $c_{1/m}(z)$ are known functions of
$z=m_c^2/m_b^2$. 
The four-quark operators 
${\cal O}_L = \bar{b}(1-\gamma_5)\gamma_\mu s
              \bar{b}(1-\gamma_5)\gamma_\mu s$ and 
${\cal O}_S = \bar{b}(1-\gamma_5) s \bar{b}(1-\gamma_5) s$ are
evaluated for $B_s$ and $\bar{B}_s$ mesons as initial and final states
respectively.
Normalizing with the total width $\Gamma_{B_s}$, one arrives at
\begin{eqnarray}
  \lefteqn{
    \left(\frac{\Delta\Gamma}{\Gamma}\right)_{B_s}
    =
    \frac{16\pi^2 B(B_s\rightarrow Xe\nu)}{
      g(z) \tilde{\eta}_{QCD}}
    \frac{f_{B_s}^2 M_{B_s}}{m_b^3} |V_{cs}|^2
    }
  & &
  \nonumber\\
  & &
  \times
  \Biggl[
    G(z) \frac{8}{3} B_B(m_b)
    + G_S(z) \frac{5}{3} \frac{B_S(m_b)}{{\cal R}(m_b)^2}
  \nonumber\\
  & &
    \;\;\;
    + \sqrt{1-4z}\,\delta_{1/m}
  \Biggl]
  \\
  & = &
  \left(
    \frac{f_{B_s}}{230\,\mbox{MeV}}
  \right)^2
  \times
  \nonumber\\
  & &
  \left[
    0.007\,B_B(m_b)
    + 0.132\,\frac{B_S(m_b)}{{\cal R}(m_b)^2}
    -0.078
  \right],
  \label{eq:formula}
\end{eqnarray}
where $B_B(m_b)$ and $B_S(m_b)$ are $B$-parameters defined as
\begin{eqnarray}
  B_B(\mu) & = &
  \frac{
    \langle\bar{B}_s|{\cal O}_L(\mu)|B_s\rangle
    }{
    \frac{8}{3} f_{B_s}^2 M_{B_s}^2
    },
  \\
  B_S(\mu) & = &
  \frac{
    \langle\bar{B}_s|{\cal O}_S(\mu)|B_s\rangle
    }{
    -\frac{5}{3} f_{B_s}^2 M_{B_s}^2
    }
  \times {\cal R}(\mu)^2,
\end{eqnarray}
and 
\begin{equation}
  {\cal R}(m_b) \equiv
  \frac{\bar{m}_b(m_b) + \bar{m}_s(m_b)}{M_{B_s}}
  = 0.81(3).
\end{equation}
is a ratio of matrix elements of heavy-light axial current and
pseudo-scalar density \cite{Hashimoto:2000yh}.
\footnote{
  The same notation ${\cal R}(m_b)$ is used for a different quantity
  in \cite{Becirevic:2000sj}.
}
Other notations may be found in \cite{Beneke:1999sy}.
We note that for $(\Delta\Gamma/\Gamma)_{B_s}$ the matrix element of
${\cal O}_S$ gives dominant contribution while
the effect of ${\cal O}_L$ is negligible.
The $1/m$ correction $\delta_{1/m}$ is expressed in terms of matrix
elements of higher dimensional $\Delta B$=2 operators, and is found to
be sizable (and destructive) by an evaluation using the vacuum
saturation approximation \cite{Beneke:1996gn}. 

To obtain the prediction for $(\Delta\Gamma/\Gamma)_{B_s}$, 
nonperturbative inputs are necessary for the decay constant $f_{B_s}$
and the $B$-parameters $B_B$ and $B_S$.
The quenched lattice calculation of the $B$ meson decay constant is
quite stable over years in the sense that results from many groups
using different lattice actions agree, and a recent summary 
is $f_{B_s}$ = 195(20) MeV \cite{Hashimoto:2000bk}.
Several groups are recently performing unquenched calculations that
include the effect of light quark loops.
Although the calculation is computationally much demanding, the
unquenched QCD simulation is being realistic on recent dedicated
machines or on commercial supercomputers.
Recent results suggest that the decay constants become higher with the
dynamical quark effect, and the most recent summary by Bernard at
Lattice 2000 is $f_{B_s}$ = 230(30) MeV \cite{Bernard:2001ki}, which
we use in this work.

\section{Lattice calculation of $B$ parameters}
Present work is an extension of previous calculations by Hiroshima
group \cite{Hashimoto:2000yh,Hashimoto:2000eh}, and preliminary
results are already reported in \cite{Yamada:2001ym,Yamada_BCP4}.

We use the lattice NRQCD action for heavy quark including all $1/M$
corrections consistently. 
Therefore, no extrapolation in the heavy quark mass is necessary in
contrast to the work by APE collaboration \cite{Becirevic:2000sj}, in
which they employ the relativistic action for relatively light heavy
quark and extrapolate the results to $m_b$.

We perform a study of systematic errors on quenched ($N_f$=0)
lattices.
We calculate the $B$-parameters in four different methods that have
different systematic errors coming from higher orders in the $1/M$ and 
$\alpha_s$ expansions, in order to see the associated systematic
uncertainties. 
\footnote{
  For a detailed description of the four methods, see
  \cite{Yamada:2001ym}. 
}
In addition, the simulations are repeated at three lattice spacings to
see the systematic error depending on lattice spacing $a$.

\begin{figure}[tb]
  \leavevmode\psfig{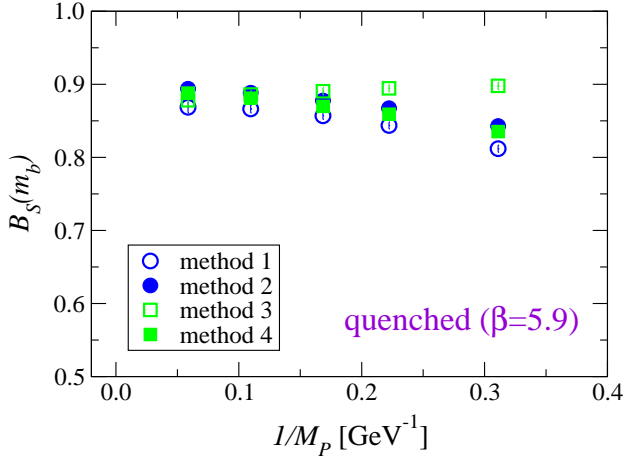}
  \caption{
    Quenched results for $B_S(m_b)$ as a function of $1/M_P$ at
    $\beta$=5.9. 
    Results with four different methods are plotted.
    }
  \label{fig:BS_b5.9}
\end{figure}

\begin{figure}[tb]
  \leavevmode\psfig{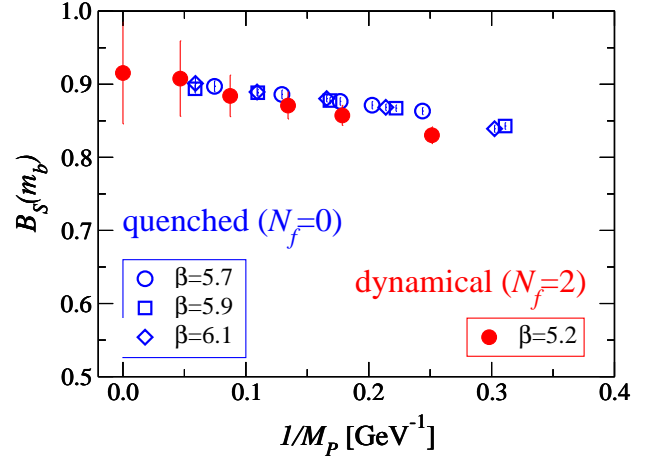}
  \caption{
    $B_S(m_b)$ as a function of $1/M_P$ at three lattice spacings.
    The method 2 is used.
    Unquenched result is also plotted (filled circles).
    }
  \label{fig:BS_all}
\end{figure}

Results for $B_S(m_b)$ are plotted in Figure~\ref{fig:BS_b5.9}, from
which we can observe that the disagreement among the different methods
becomes larger for larger $1/M$.
That behavior is expected, because the $1/M$ expansion is truncated at
first order in our calculation.
Nevertheless the result at the $B_s$ meson mass
($1/M_P\simeq$~0.19~GeV$^{-1}$) is under reasonable control.
A similar plot is shown in Figure~\ref{fig:BS_all} for three lattice
spacings in the quenched approximation.
We find a good agreement among the results from different lattices,
which indicates that the systematic error associated with the lattice
discretization is well controlled.

At this conference, we presented a new result from unquenched
simulations.
As one of the major projects of the JLQCD collaboration, we are
performing a two-flavor (dynamical $u$ and $d$ quarks) QCD 
simulation with a nonperturbatively $O(a)$-improved quark at
$\beta$=5.2, $c_{\mathrm{sw}}$=2.02 on a 20$^3\times$48 lattice, which
corresponds to the lattice spacing $a\simeq$ 0.1~fm.
A preliminary report of this simulation may be found in
\cite{Aoki:2001yi}.
On this lattice, we are calculating the $B$-parameters and a very
preliminary result is plotted in Figure~\ref{fig:BS_all}.
Although the result is slightly lower than the quenched data, the
difference is statistically not significant, and our conclusion at
this stage is that the effect of dynamical quarks is not sizable for
the $B$-parameters.

Our (preliminary) results for $B$-parameters in the quenched ($N_f$=0)
and unquenched ($N_f$=2) calculations are 
\begin{eqnarray}
  B_B(m_b)
  & = &
  \left\{
    \begin{array}[c]{ll}
      0.85(2)(8) & (N_f = 0)\\
      0.83(3)(8) & (N_f = 2)
    \end{array}
  \right.,
  \\
  B_S(m_b)
  & = &
  \left\{
    \begin{array}[c]{ll}
      0.87(1)(9) & (N_f = 0)\\
      0.84(6)(8) & (N_f = 2)
    \end{array}
  \right..
\end{eqnarray}
The systematic errors are evaluated using the variation among four
different methods as a guide.
Using (\ref{eq:formula}) the result for the width difference is
obtained as
\begin{equation}
  \left(
    \frac{\Delta\Gamma}{\Gamma}
  \right)_{B_s}
  =  0.097^{+0.014}_{-0.035} \pm 0.025 \pm 0.020 \pm 0.016,
  \label{eq:result}
\end{equation}
where errors are from residual scale $\mu$ dependence remaining in the
calculation of $G(z)$ and $G_S(z)$ in (\ref{eq:formula}), $f_{B_s}$,
$B_S$, and the uncertainty in the $1/m$ corrections, respectively.
The central value 0.097 is somewhat smaller than our previous estimate 
0.107(26)(14)(17) \cite{Hashimoto:2000eh} due to a smaller input for
$f_{B_s}$, which was previously 245(30)~MeV.
The effect of unquenching is less significant ($-6\%$).
The experimental results from LEP and SLD were summarized by Boix at
this conference \cite{Boix_BCP4} as 
$(\Delta\Gamma/\Gamma)_{B_s} = 0.16^{+0.08}_{-0.09}$.

One of the major systematic errors in (\ref{eq:result}) comes from the
uncertainty in $f_{B_s}$, which may be avoided by considering a ratio
$(\Delta\Gamma/\Delta M)_{B_s}$, once $\Delta M_s$ is measured
\cite{Beneke:1999sy}. 
We obtain
\begin{eqnarray}
  \lefteqn{
    \left(\frac{\Delta\Gamma}{\Delta M}\right)_{B_s}
    =
    \frac{\pi}{2}\,\frac{m_b^2}{M_W^2}
    \left|
      \frac{V_{cb}^* V_{cs}}{V_{tb}^* V_{ts}}
    \right|^2
    \frac{1}{\eta_B(m_b) S_0(x_t)}
    }
  \nonumber\\
  & & \times
  \left[
    \frac{8}{3}G(z)
    + \frac{5}{3} G_S(z) 
    \frac{B_S(m_b)}{B_B(m_b)}\,
    \frac{1}{{\cal R}(m_b)^2}
  \right.
  \nonumber\\
  & & \;\;\;\;
  \left.
    + \frac{\sqrt{1-4z}\,\delta_{1/m}}{B_B(m_b)}
  \right],
  \nonumber\\
  & = &
  \left(
    0.20 + 6.00\,\frac{B_S(m_b)}{B_B(m_b)} - 2.85
  \right)
  \times 10^{-3}
  \nonumber\\
  & = &
  ( 3.5 ^{+0.4}_{-1.3} \pm 0.6 \pm 0.6 )\times 10^{-3},
\end{eqnarray}
where errors come from $\mu$, $B_S/B_B$, and the $1/m$ corrections,
respectively.

\section{Comparison with other approaches}
There are two other approaches to calculate the $B$ meson
$B$-parameters on the lattice.

One is the HQET approach, in which the $b$ quark is treated as a
static color source, and it is a naive limit $M\rightarrow\infty$ of
the NRQCD action.
Gim\'enez and Reyes calculated $B_B$ and $B_S$ in this method
\cite{Gimenez:2001en,Gimenez:2001jj}, in which an unquenched
calculation is also performed (but with an unimproved action).

The other is the relativistic approach, in which the relativistic
lattice action is used for heavy quark.
In order to avoid large discretization error growing as a power of
$aM$, one cannot simulate the $b$ quark and has to treat relatively
light heavy quark with mass around 1--2~GeV.
An extrapolation in the heavy quark mass is then attempted.
The crucial point for this method is to find a heavy quark
mass region where $M$ is small enough so that the discretization error
of $O((aM)^2)$ is under control, and at the same time $M$ is large
enough to justify the use of heavy quark expansion in the
extrapolation. 
In general it may not be easy to identify such a region with
confidence, and the applicability of the method should be studied
carefully depending on the quantities to calculate.
Calculation of the $B$-parameters in this method has been presented 
by APE \cite{Becirevic:2000sj} and UKQCD \cite{Flynn:2000hx}
collaborations.

\begin{figure}[tb]
  \leavevmode\psfig{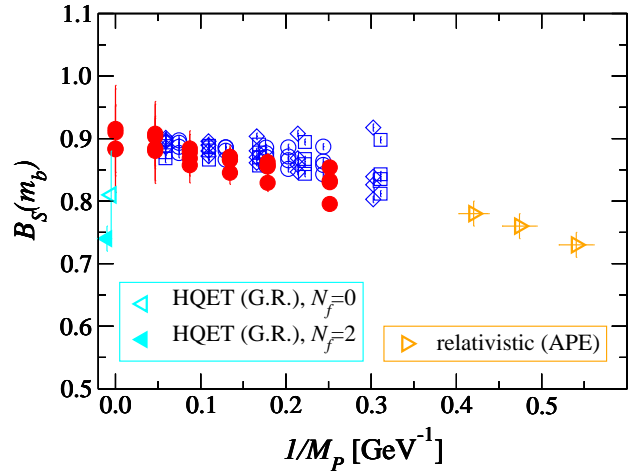}
  \caption{
    $B_S(m_b)$ from different approaches.
    Our results are shown by open symbols ($N_f$=0) and solid circles
    ($N_f$=2). 
    }
  \label{fig:BS_comp}
\end{figure}

In Figure~\ref{fig:BS_comp} we compare our results with other
approaches.
While our data cover the physical region 
$1/M_P$ = 0.05--0.3~GeV$^{-1}$, the HQET results are plotted at the
static limit $1/M_P$ = 0, and the relativistic results are shown in
the charm quark mass region $1/M_P$ = 0.4--0.5~GeV$^{-1}$.
Although we did not attempt to fit the whole data with a single curve,
it seems that the results of different approaches agree reasonably
well. 
It suggests that the lattice calculation of $B_S$ is in good shape,
even though the individual method might have its own systematic
uncertainty. 

Besides the calculation of $B$-parameters, the APE collaboration
proposed to express $\Delta\Gamma_{B_s}$ using $\Delta M_d$ as a
normalization \cite{Becirevic:2000sj}
\begin{eqnarray}
  \left(\frac{\Delta\Gamma}{\Gamma}\right)_{B_s}
  & = &
  K \left(
    \tau_{B_s}\,\Delta M_d \frac{M_{B_s}}{M_{B_d}}
  \right)
  \left(
    \frac{f_{B_s}^2 \hat{B}_{B_s}}{f_{B_d}^2 \hat{B}_{B_d}}
  \right)
  \,
  \left|
    \frac{V_{ts}}{V_{td}}
  \right|^2
  \nonumber\\
  & & \times
  \left[
    G(z) 
    + G_S(z) \frac{5}{8} 
    \frac{B_S(m_b)}{B_B(m_b)} \frac{1}{{\cal R}(m_b)^2}
  \right.
  \nonumber\\
  & & \;\;\;\;
  \left.
    + \frac{3}{8}
    \frac{\sqrt{1-4z}\,\delta_{1/m}}{B_B(m_b)}
  \right],
  \label{eq:formula_APE}
\end{eqnarray}
and quoted a much smaller result 
$0.047 \pm 0.015 \pm 0.016$.
$K$ is a known factor, and an experimental value may be used for
$(\tau_{B_s}\,\Delta M_d \frac{M_{B_s}}{M_{B_d}})$.
In the ratio 
$\frac{f_{B_s}^2 \hat{B}_{B_s}}{f_{B_d}^2 \hat{B}_{B_d}}$
the bulk of systematic error should cancel in the lattice calculation,
and the result has much smaller uncertainty.
The problem in this normalization is, however, that the large
uncertainty enters implicitly through the (ratio of) CKM elements
$|V_{ts}/V_{td}|^2$, for which they use a value obtained from a global
fit of the CKM elements.
We do not take this strategy, because the several sources of
(theoretical) systematic errors are hidden in the CKM fit, and it
makes the estimate of systematic uncertainties less transparent.

\section{Requirements for further improvement}
In order to improve the accuracy of the prediction (\ref{eq:result}),
better determination of $f_{B_s}$ is important.
For the unquenched lattice calculation, more work is necessary to
reach the situation in the quenched case, where consistency is checked
among several different approaches by many groups.
The similar work is necessary for the $B$-parameters, for which only a
few unquenched calculations are available including ours.

In (\ref{eq:formula}), it is evident that the $1/m$ correction gives a
large negative contribution to the width difference.
Therefore, its reliable evaluation is quite important.
At present, it is estimated using the vacuum saturation approximation,
and thus nonperturbative calculation on the lattice is desirable.
Although it is hard to determine the (perturbative) matching
coefficients for higher dimensional operators, the lattice calculation
should still be useful to have an idea how reliably the $1/m$
corrections are estimated. 

\section{Conclusions}
The JLQCD collaboration started a lattice calculation of $B_B$ and
$B_S$ including the effect of sea quarks.
A preliminary result shows that the quenching effect is not
substantial for these quantities.

We show that the calculations with several different methods on three
lattice spacings show a reasonable agreement in the quenched
approximation. 
It indicates that, using the lattice NRQCD, the systematic errors are
under control. 

The results for $(\Delta\Gamma/\Gamma)_{B_s}$ still have large
uncertainty.
Better estimation of the nonperturbative inputs as well as the $1/m$
corrections will be necessary to improve the accuracy.

\section*{Acknowledgments}
We thank the members of JLQCD collaboration including
Ken-Ichi~Ishikawa and Tetsuya~Onogi for discussions.
This work is supported by the Supercomputer Project No. 54 (FY2000) of
High Energy Accelerator Research Organization (KEK), also in part by
the Grant-in-Aid of the Ministry of Education (No. 11740162).
N.Y. is supported by the JSPS Research Fellowship.


\section*{References}
\begin{thebibliography}{99}

\bibitem{Dunietz:2000cr}
  See, for example,
  I.~Dunietz, R.~Fleischer and U.~Nierste,
  hep-ph/0012219,
  and references therein.

\bibitem{Beneke:1996gn}
  M.~Beneke, G.~Buchalla and I.~Dunietz,
  Phys.\ Rev.\ D {\bf 54}, 4419 (1996).

\bibitem{Beneke:1999sy}
  M.~Beneke, G.~Buchalla, C.~Greub, A.~Lenz and U.~Nierste,
  Phys.\ Lett.\ B {\bf 459}, 631 (1999).

\bibitem{Beneke:2000cu}
  M.~Beneke and A.~Lenz,
  talk given at UK Phenomenology Workshop on Heavy Flavor and CP
  Violation, Durham, England, 17-22 Sep 2000,
  hep-ph/0012222.

\bibitem{Hashimoto:2000yh}
  S.~Hashimoto, K.-I.~Ishikawa, T.~Onogi and N.~Yamada,
  Phys.\ Rev.\ D {\bf 62}, 034504 (2000).

\bibitem{Hashimoto:2000eh}
  S.~Hashimoto, K.-I.~Ishikawa, T.~Onogi, M.~Sakamoto, N.~Tsutsui and
  N.~Yamada, 
  Phys.\ Rev.\ D {\bf 62}, 114502 (2000).

\bibitem{Yamada:2001ym}
  N.~Yamada {\it et al.} [JLQCD Collaboration],
  Nucl.\ Phys.\ B (Proc.\ Suppl.)\ {\bf 94}, 379 (2001).

\bibitem{Becirevic:2000sj}
  D.~Becirevic, D.~Meloni, A.~Retico, V.~Gimenez, V.~Lubicz and
  G.~Martinelli, 
  Eur.\ Phys.\ J.\ C {\bf 18}, 157 (2000).

\bibitem{Hashimoto:2000bk}
  S.~Hashimoto,
  Nucl.\ Phys.\ B (Proc.\ Suppl.)\ {\bf 83}, 3 (2000).

\bibitem{Bernard:2001ki}
  C.~Bernard,
  Nucl.\ Phys.\ B (Proc.\ Suppl.)\ {\bf 94}, 159 (2001).

\bibitem{Yamada_BCP4}
  N.~Yamada, 
  presented at the BCP4 conference, to appear in the proceedings. 

\bibitem{Aoki:2001yi}
  S.~Aoki {\it et al.}  [JLQCD Collaboration],
  Nucl.\ Phys.\ B (Proc.\ Suppl.)\ {\bf 94}, 233 (2001).

\bibitem{Boix_BCP4}
  G.~Boix, 
  presented at the BCP4 conference, to appear in the proceedings. 

\bibitem{Gimenez:2001en}
  V.~Gimenez and J.~Reyes,
  Nucl.\ Phys.\ B (Proc.\ Suppl.)\ {\bf 93}, 95 (2001).

\bibitem{Gimenez:2001jj}
  V.~Gimenez and J.~Reyes,
  Nucl.\ Phys.\ B (Proc.\ Suppl.)\ {\bf 94}, 350 (2001).

\bibitem{Flynn:2000hx}
  J.~Flynn and C.~J.~Lin,
  talk given at UK Phenomenology Workshop on Heavy Flavor and CP
  Violation, Durham, England, 17-22 Sep 2000,
  hep-ph/0012154. 

\end{thebibliography}
\end{document}